\documentclass[a4paper,fleqn,12pt]{article}
\usepackage{color}
\usepackage{fancyhdr}
\usepackage{graphicx}
\usepackage{amsmath}
\usepackage{amssymb}
\usepackage{array}
\usepackage{multicol}
\definecolor{mark}{rgb}{0.,0.,0.2}
\title{\vspace{-10mm}{\bf \boldmath 
First Beam Observation and Near Future Plans at SPring-8 LEPS2 Experiments}}
\author{N.~Muramatsu for the LEPS2 Collaboration\\
        {\small \it Research Center for Electron Photon Science, Tohoku University} \\
        {\small \it 1-2-1 Mikamine, Taihaku, Sendai, Miyagi 982-0826, Japan}}
\date{Presented at BARYONS 2013 on June 27, 2013}
\makeatletter

\def\@makefnmark{{\bf \textsuperscript{[\@thefnmark]}}}
\makeatother
\begin{document}
\maketitle
\thispagestyle{fancy}
\rhead{\vspace*{5mm} ELPH Report 2044-13 \hspace{5mm} July, 2013}
\renewcommand{\headrulewidth}{0pt}
\begin{abstract}
The first photon beam was successfully produced by laser Compton backscattering at 
the LEPS2 beamline, which was newly constructed at SPring-8 for the purpose to increase 
the beam intensity one order of magnitude more than that of the LEPS experiments and 
to achieve the large acceptance coverage with high resolution detectors. The BGOegg 
electromagnetic calorimeter with associated detectors are being set up at the LEPS2 
experimental building for the physics programs, including the searches for $\eta$'-bound 
nuclei and highly excited baryon resonances. In parallel to the BGOegg experiments, 
the LEPS2 charged particle spectrometer will be prepared inside the 1~Tesla solenoidal 
magnet, transported from the BNL-E949 experiment.
\end{abstract}

\begin{multicols}{2}

\section{From LEPS to LEPS2}

At the LEPS experiments, which started from 1999, the ultraviolet (UV) laser light with 
the wavelength of 355~nm has been injected into the 8~GeV electron storage ring in order 
to produce the laser Compton backscattering (LCS) photon beam \cite{muramatsu}. The beam 
intensity has exceeded $\sim10^6$ Hz in the tagged photon energy range of 1.5--2.4~GeV. 
High polarization is transferred from the laser light to the photon beam, so that the 
t-channel exchange reaction with a forward meson photoproduction is usable as a parity 
filter. Such reactions have been extensively investigated by the LEPS forward spectrometer, 
which covers the polar angle region within $\pm$20$^\circ$ and $\pm$10$^\circ$ in the 
horizontal and vertical directions, respectively. For example, we have analyzed the 
forward K$^{*0}$ $\to$ K$^+$$\pi^-$ production from a proton target with the identification
of the associated $\Sigma^+$ in the missing mass distribution \cite{hwang}. In this study,
the LCS photon beam energy was upgraded up to 2.9~GeV by using a deep UV laser. The dominance 
of natural parity exchange in the t-channel was observed based on the parity spin asymmetry 
measurement, indicating the evidence of the controversial scalar meson $\kappa$(800).

On the other hand, the further systematic studies of hadron photoproduction generally
require larger acceptance coverage and higher photon beam intensity. Recently, we measured 
the differential cross sections of backward $\eta$' photoproduction with the forward proton 
detection at the LEPS forward spectrometer \cite{morino}. It was suggested that the bump 
structure at W$\sim$2.3~GeV may be enhanced in the most backward angles, which are not 
covered by the CLAS experiments \cite{clas}. There may be a resonance contribution with 
a high angular momentum, but it is needed to increase the statistics possibly with a wide 
angular acceptance. Because of the demands represented by this measurement, we are motivated 
to start the LEPS2 project at another beamline of SPring-8 \cite{yosoi}.

\section{LEPS2 Facility}

\begin{figure*}[t]
 \centering
 \includegraphics[width=16cm]{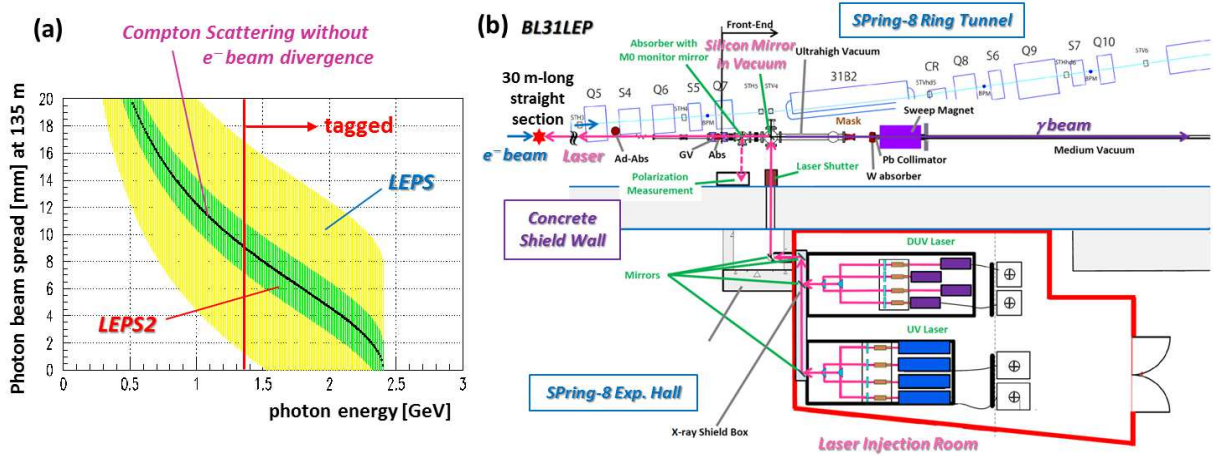}
 \caption{(a) Photon beam spread at the 135~m downstream from the Compton scattering point.
          The solid line comes from the purely kinematical calculation of backward Compton
          scattering. Actually, it is smeared by the electron beam divergences at the LEPS
          and LEPS2 beamlines, as shown in the shaded areas (corresponding to RMS). (b) The 
          schematic view of the LEPS2 facility around the newly installed vacuum chambers 
          and the laser side-injection system.}
 \label{fig:leps2bl}
\end{figure*}

The LEPS2 beamline utilizes a 30~m-long straight section, the number of which is limited 
to only 4 of the 62 beamlines at SPring-8. The horizontal divergence of the electron beam 
at this straight section is reduced to 14~$\mu$rad in $\sigma$, while the usual 7.8~m-long 
straight section including the LEPS beamline provides the divergence of 58~$\mu$rad. 
As shown in Fig.~\ref{fig:leps2bl}(a), the LCS photon beam spread at the LEPS2 beamline 
is determined not largely by the electron beam divergence but mostly by the Compton 
scattering angle, which is calculated by the kinematics depending on the photon energy. 
This achieves a well collimated photon beam with the radius below 10~mm even at the 135~m 
downstream from the Compton scattering point, enabling us to construct a large experimental 
site outside the storage ring building.

We also aim to increase the tagged photon beam intensity nearly up to 10$^7$ Hz for the photon 
energy range below 2.4~GeV by using the UV lasers with the wavelength of 355~nm. We plan 
the simultaneous injection of four lasers, whose output power have increased from 8~W to 16~W 
or 24~W. The multiple laser operation has become possible thanks to the reduced electric power 
consumption of those new lasers. For the simultaneous injection, we have installed new beamline 
chambers with large apertures \cite{yorita}.
As shown in Fig.~\ref{fig:leps2bl}(b), the laser injection from 
the side concrete shield wall into the SPring-8 ring tunnel is adopted in order to reduce the 
total length and apertures of the newly installed chambers. The radiation level at the laser 
injection room is also lowered with the minimized shield materials, so that we can easily 
access the laser optics for adjustments. The large mirror to guide the four laser beams from 
the side direction toward the straight section inside a vacuum chamber has been modified to 
possess a horizontal slit and a small center hole for avoiding the heat input from x-rays and 
the e$^+$e$^-$ conversions of the LCS photon beam, respectively. In addition to the LCS beam 
with the UV lasers, we plan to improve the intensity of a high energy photon beam using the 
deep UV lasers with the wavelength of 266~nm. We expect the tagged photon intensity approaching 
to 10$^6$ Hz with the techniques similar to the UV case.

\begin{figure*}[t]
 \centering
 \includegraphics[width=16cm]{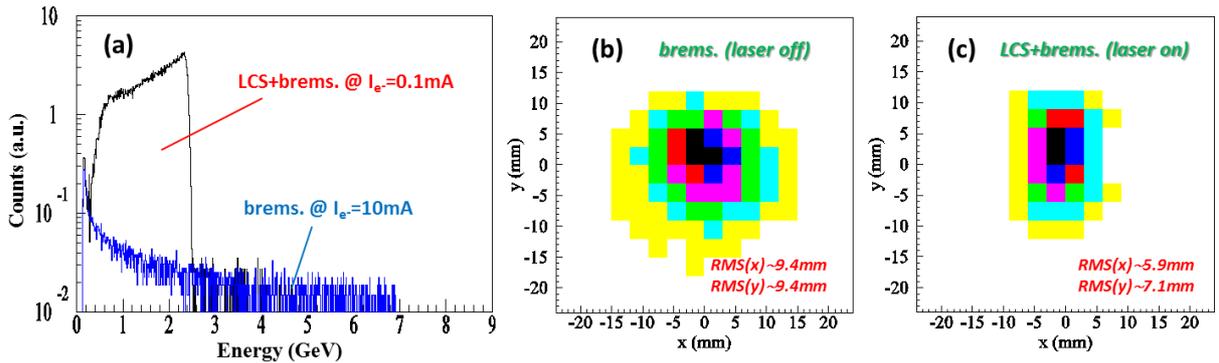}
 \caption{(a) Photon beam energy spectra with and without laser injection at the LEPS2
          beamline. (b) and (c) Photon beam profiles at the LEPS2 experimental site
          without and with laser injection, respectively.}
 \label{fig:bmprop}
\end{figure*}

The construction of the LEPS2 beamline was approved in 2010. The LEPS2 experimental site, whose 
size is 12~m $\times$ 18m in area with the height of 10~m, was built in 2011. The constructed 
experimental building has the volume 15 times larger than the experimental hutch of the LEPS 
experiments. In 2011, we also transported the BNL-E949 magnet, which is a 1~Tesla solenoidal 
magnet with the weight of 400 tons, and successfully installed it into the LEPS2 experimental 
building. This magnet has the bore diameter of 2.96~m and the inner length of 2.22~m, where 
the charged particle spectrometer described later will be placed. The beamline vacuum chambers 
and the laser injection system were finally installed and aligned in late 2012. 

\section{First Beam Observation}

After adjusting the electron beam orbit at the straight section and degassing the absorbers
and the mask with synchrotron radiation, the first LCS photon beam at the LEPS2 beamline was 
produced with the injection of a single UV laser beam during the machine study on 27~Jan.~2013. 
The simultaneous injection of three UV laser beams was then tested on 2~Apr.~for the further 
investigation of beam properties as described below. Figure~\ref{fig:bmprop}(a) shows the photon 
energy spectra, which have been measured by a large Bi$_4$Ge$_3$O$_{12}$ (BGO) crystal with 
the diameter of 8~cm and the length of 30~cm. The blue histogram represents the spectrum for 
a bremsstrahlung photon beam without laser injection at the electron beam current of 10~mA, 
while the black histogram shows that for a LCS photon beam with the contamination of 
a bremsstrahlung beam at the current of 0.1~mA. The two spectra are normalized by the electron 
beam current, the data taking time, and the DAQ live time. The lowest photon energy part of 
the LCS beam spectrum is cut off at $\sim$0.65~GeV because of a $\phi$7~mm lead collimator 
inside the storage ring tunnel. The intensity of a LCS photon beam, corresponding to the 
spectrum height, is nearly two orders of magnitude higher than that of a bremsstrahlung beam, 
although the straight section is long.

The photon beam profile was measured by the 16~ch.~$\times$ 16~ch.~two-dimensional array of 
3~mm-square scintillating fibers with a 0.5~mm-thick aluminum converter, which was placed 
between the plastic scintillators for the charge veto and the trigger. We refer to this 
detector as a beam profile monitor (BPM). As shown in Fig.~\ref{fig:bmprop}(b), the profile 
of a bremsstrahlung photon beam is a bit spread with the RMS radius of $\sim$9~mm because 
of the electron beam orbit structure at the 30~m-long straight section. On the other hand, 
the profile with laser injection is well collimated with the dominance of a LCS photon beam
(Fig.~\ref{fig:bmprop}(c)), resulting in the RMS radius of $\sim$6~mm as expected from 
Fig.~\ref{fig:leps2bl}(a).

\begin{figure*}[t]
 \centering
 \includegraphics[width=16cm]{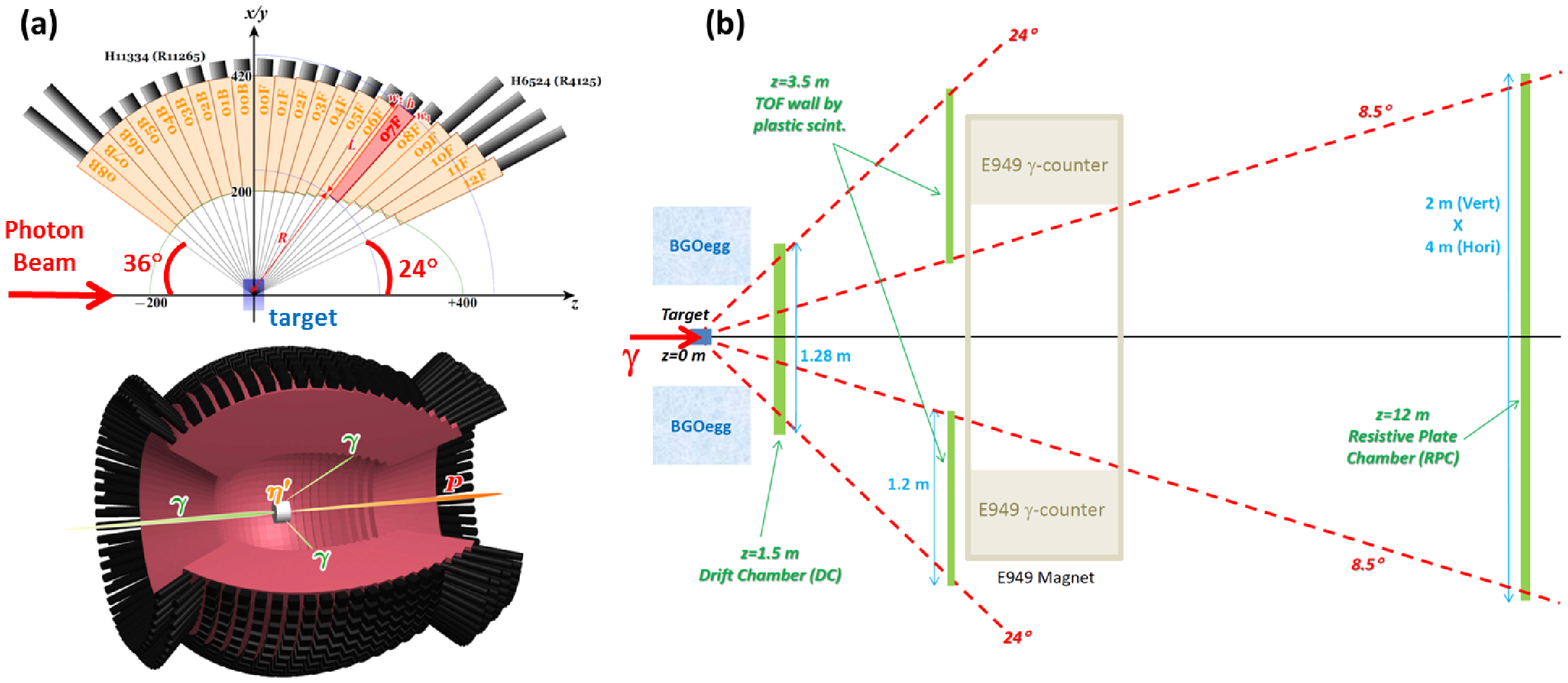}
 \caption{(a) Side views of the BGOegg electromagnetic calorimeters. (b) A schmatic top view
          of the forward detector setup for charged particles.}
 \label{fig:bgoegg_setup}
\end{figure*}

The photon beam intensity was measured with the simultaneous three laser injection. 
The intensity is usually measured by counting the recoil electron rate at the tagger, but 
its optimization is still underway. Therefore, we adopted the following two alternative 
methods for the intensity measurement: One of them is the estimation from the change of 
electron beam life before and after the laser injection. The electron beam life corresponding 
to the Compton scattering loss was measured to be $\sim$120 hours. This measurement was 
performed with the electron beam current of 100~mA, so the backward Compton scattering rate 
corresponding to the photon beam intensity was calculated to be $\sim$7~MHz for 
0$<$E$_\gamma$$<$2.4~GeV. Another estimation was done by the measurement of the e$^+$e$^-$ 
conversion rate at the BPM. The BPM trigger rates were 29.0 kHz and 3.7 kHz for the cases 
with laser on and off, respectively. By taking into account the pair production cross sections 
at the converter and the trigger scintillator, the photon beam transmission through the beamline 
materials ($\sim$76\%), and the cutoff of low energy photons by the collimator, the intensity 
of a LCS photon beam is calculated to be $\sim$7~MHz, which is consistent with the estimation 
from the electron beam life. The obtained beam intensity is nearly half of our expectation, 
so that we are trying to fix the problems of laser injection optics.

\section{Experimental Programs}

Two different detector setups are being prepared for the experimental programs at the LEPS2 
facility. The first experiments will be carried out with the electromagnetic calorimeter 
called 'BGOegg' from the latter half of FY2013. The calorimeter system has been installed 
in the upstream part of the LEPS2 experimental building, and covered by a thermostatic booth 
to avoid the calorimeter gain change which is expected at the level of $-$1.5\%/$^\circ$C. 
In the center part of the LEPS2 building, the BNL-E949 solenoidal magnet has been set up 
with a pit hole. We will keep constructing the LEPS2 charged particle spectrometer inside 
the solenoid during the BGOegg experiments in parallel. The experiments with this spectrometer 
will follow the BGOegg experiments in a few years.

\subsection{BGOegg Experiments}

\begin{figure*}[t]
 \centering
 \includegraphics[width=16cm]{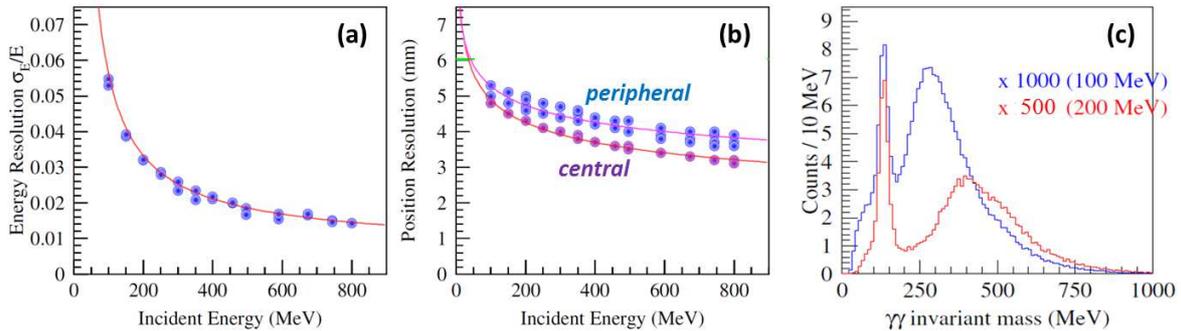}
 \caption{(a) Energy resolutions and (b) position resolutions as a function of the incident
          energy at the positron beam test of the prototype BGOegg calorimeter. The data
          points indicated as `central' in (b) were measured by the energy weights in 
          3$\times$3 BGO crystals with the beam injection into the central crystal. (c) The 
          distribution of 2$\gamma$ invariant mass measured at the partial operation (only 
          forward 300 channels) of the BGOegg calorimeter. The two histograms, which must 
          be scaled by the indicated magnification factors, are plotted for the different 
          threshold values of an individual $\gamma$ energy.}
 \label{fig:bgoegg_perform}
\end{figure*}

The BGOegg electromagnetic calorimeter is an 'egg'-shaped assembly of 1320 BGO crystals, 
as shown in Fig.~\ref{fig:bgoegg_setup}(a). Each crystal with the shape of a truncated square 
pyramid has the longitudinal length of $\sim$220~mm, corresponding to 20 radiation length. 
The BGO crystals are separately stacked in the forward 13 layers and the backward 9 layers 
of the 60 crystal ring without partition supports. The BGOegg covers the large geometrical 
acceptance from 24$^\circ$ to 144$^\circ$ in polar angles. We will install a cylindrical 
drift chamber (CDC) and plastic scintillators inside the BGOegg for the detection of charged 
particles. The BGOegg calorimeter was constructed at Research Center for Electron Photon 
Science (ELPH), Tohoku University, and was successfully transported to SPring-8 in December
2012.

The performance of the BGOegg calorimeter has been examined at ELPH by injecting a positron 
beam into a 5$\times$5 crystal array with the front coverage by scintillating fiber hodoscopes 
\cite{ishikawa}. As shown in Fig.~\ref{fig:bgoegg_perform}(a), the measured energy resolution 
varies as a function of the incident energy, resulting in $\Delta$E/E$=$1.3\% at 1~GeV. The 
positron beam position was also measured from the shower distribution in 3$\times$3 crystals, 
and was compared with the hodoscope information. The position resolution was estimated to be 
3.1~mm at the center part of the crystal array for the incident energy of 1~GeV. (See 
Fig.~\ref{fig:bgoegg_perform}(b).)

In January 2013, the BGOegg was partially operated at the LEPS2 beamline for a commissioning 
purpose. Photomultipliers, high voltage suppliers, and a related DAQ system were prepared 
only for the most forward 5 layers or 300 BGO crystals. A test beam of the LCS photons was 
irradiated onto a 20~mm-thick carbon target at the BGOegg center. Shower activities which 
coincide with the tagger signal timing were clustered after the rough gain calibrations of 
300 crystals. In Fig.~\ref{fig:bgoegg_perform}(c), two of such clusters were combined to 
calculate the invariant mass with the different conditions of energy thresholds. A clear peak 
of the $\pi^0 \to \gamma\gamma$ decays is observed with the mass resolution of $\sim$15~MeV, 
which is slightly worse than the expected resolution because the precise energy calibrations 
has not been done. The $\eta \to \gamma\gamma$ decays require larger opening angles, so the 
corresponding signal peak is not clear in this plot due to the limited geometrical acceptance.

\subsection{Physics Programs expected at the BGOegg Experiments}

\begin{figure*}[t]
 \centering
 \includegraphics[width=16cm]{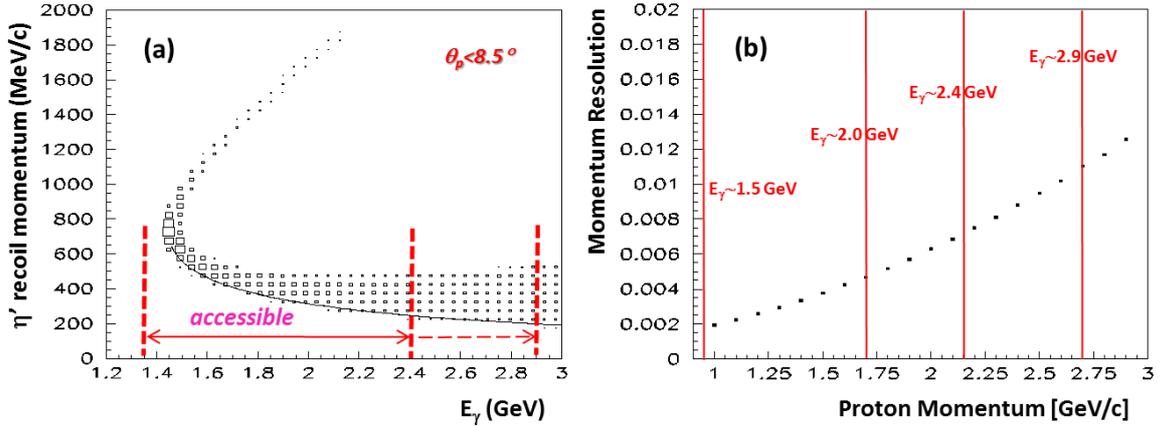}
 \caption{(a) The simulated $\eta$' recoil momenta as a function of the photon beam energy
          in the case that the final state proton in the $\gamma$p $\to$ $\eta$'p reaction
          is detected at the RPC, covering the extremely forward angles less than 8.5$^\circ$. 
          The photon energy ranges accessible with the UV and deep UV lasers at the LEPS2 
          experiments are also indicated. (b) The expected momentum resolutions as a function 
          of the proton momentum, measured from the TOF at the RPC. The maximum proton momenta 
          at the various photon energies are also indicated.}
 \label{fig:bgoegg_phys}
\end{figure*}

We are considering to search for the $\eta$'-bound nuclei as one of main physics programs 
at the BGOegg experiments. By the partial restoration of the chiral symmetry, the $\eta$' 
mass is theoretically expected to decrease by $\sim$150~MeV at a normal nuclear density 
\cite{nagahiro}. This produces a potential for the $\eta$' meson to be bound inside 
a nucleus. Such a state may be accessible by injecting a high energy photon into 
a nucleus and by striking a proton to the extremely forward direction.

At the BGOegg experiments, the forward acceptance hole of the calorimeter is covered by 
TOF detectors to measure the momenta of charged particles, as shown in 
Fig.~\ref{fig:bgoegg_setup}(b). For the extremely forward region at $\theta$$<$8.5$^\circ$, 
a 2~m$\times$4~m wall of resistive plate chambers (RPC), whose individual size is 
20~cm$\times$100~cm, will be placed at the 12~m downstream of the target. The time resolution 
of the RPC was measured by a beam test at the LEPS beamline, and was estimated to be 50 psec 
\cite{tomida}. We are planning to cover the forward angles between the RPC wall and 
the calorimeter (8.5$^\circ$$<$$\theta$$<$24$^\circ$) by the TOF wall of plastic scintillators 
at the just upstream of the E949 solenoid. The positions and angles of forwardly produced 
charged particles will be measured by a drift chamber (DC), which has an effective diameter 
of 1280~mm. The DC consists of 6 planes with three different wire angles, possessing 
16~mm-square drift cells. The position resolution of 130 $\mu$m has been confirmed at a beam 
test \cite{hashimoto}.

The setup of the BGOegg calorimeter with the forward charge detection is suitable for 
the $\eta$'-bound nuclei search. In the case that the forward proton is detected at the RPC, 
the recoiled $\eta$' has a low momentum ($\sim$400~MeV/c) to be bound at a proton hole. (See
Fig.~\ref{fig:bgoegg_phys}(a).) The higher photon beam energy with the injection of deep UV 
laser reduces the lower limit of the recoil momentum further. The momentum resolution for 
the forward proton detected at the RPC is estimated to be less than 1\% as shown in 
Fig.~\ref{fig:bgoegg_phys}(b). This results in the missing mass resolution of 15~MeV at 
E$_\gamma$$=$2.4~GeV. The BGOegg calorimeter will be used to improve the S/N ratio, and 
it may be possibly used to detect $\eta$' decays for a line shape analysis. The $\eta$' mass 
resolution by two gamma invariant mass is estimated to be 30~MeV at E$_\gamma$$=$2.4~GeV 
without a kinematical fitting.

\begin{figure*}[t]
 \centering
 \includegraphics[width=7cm]{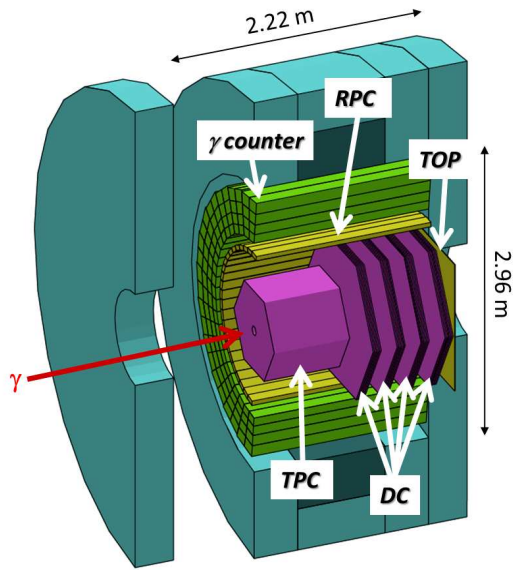}
 \caption{A schematic view of the LEPS2 charged particle spectrometer.}
 \label{fig:leps2spect}
\end{figure*}

The $\eta$'-related physics will be also explored extensively by preparing a liquid hydrogen 
or deuterium target with a long nose. For example, it may be possible to measure the $\eta$'p 
scattering length. This measurement has become possible by extending the lower limit of the 
tagged photon energy range to 1.35~GeV with the modification of a LEPS2 beamline chamber and 
by covering the kinematical region well below the $\eta$' production threshold of E$_\gamma 
\sim$1.45~GeV. Our detector setup has the feasibility for this study because, at low photon 
energies, a proton is directed to the extremely forward angles and the $\eta$' angle dependence 
of BGOegg acceptance gets small. At E$_\gamma$$<$2.4~GeV, the $\eta$' yield of the elementary 
photoproduction process is estimated to be $\sim$60,000 events $\times$ Branching Fraction 
$\times$ Acceptance per day by assuming the total cross section of 0.8~$\mu$b, the beam 
intensity of 10$^7$ Hz, and the liquid hydrogen target length of 4~cm.

Highly excited baryon resonances, which are not well explained by the constituent quark 
model, strongly couple to the final state of multi-mesons and a nucleon. We therefore 
plan to investigate the double meson photoproduction in the $\pi\pi$N, $\pi\eta$N, and 
$\eta\eta$N channels. Especially, the $\eta\eta$N channel works as an isospin filter, 
which only allows a resonance with I=1/2. Similar studies with partial wave analyses 
can be done at the CBELSA/TAPS experiments through the various combinations of beam and 
target polarizations. The BGOegg experiments would be able to contribute to this field 
with the high linear polarization beam at high energies, while the linear polarization 
of the CBELSA/TAPS experiments is achievable up to E$_\gamma \sim$1.7~GeV.

\subsection{LEPS2 Charged Particle Spectrometer}

After a few years run with the BGOegg detector, the next experiments with the LEPS2 charged 
particle spectrometer are planned by covering large solid angles, as shown in 
Fig.~\ref{fig:leps2spect}. The BNL-E949 magnet provides the solenoidal field of 1~Tesla, 
where we are going to place a time projection chamber (TPC) and DCs as a tracking system. 
The momentum resolution of about 1\% is expected for the polar angle region greater than 
10$^\circ$. Particle identification will be done by a cylindrical side wall of RPCs and 
a forward wall of TOP counters. A part of the DCs and the RPCs will be recycled from the 
BGOegg experiments. The most outer part of the barrel region is occupied by the sampling 
electromagnetic calorimeter, which has been originally used at the E949 experiment, although 
the energy resolution is expected to be around 10\%.

As one of the physics programs with the LEPS2 charged particle spectrometer, we will perform 
the systematic study of the pentaquark candidate $\Theta^+$, which has not been established 
yet. The reaction of $\gamma$n $\to$ K$^-\Theta^+$; $\Theta^+$ $\to$ K$_S^0$p; K$_S^0$ $\to$ 
$\pi^+\pi^-$ will be examined because the final state of this mode includes only charged 
particles. The mass resolution for the $\Theta^+$, which will be reconstructed by 
the $\pi^+\pi^-$p invariant mass, is expected to be 6~MeV. On the other hand, we are 
finalizing the exclusive analysis of the reaction $\gamma$n $\to$ K$^-\Theta^+$; $\Theta^+$ 
$\to$ K$^+$n in the LEPS data by taking into account the importance of the final state nucleon
identification \cite{fb20}. The new data to confirm the observed structure in the K$^+$n mass 
spectrum is being accumulated with the LEPS forward spectrometer, whose trigger counter has 
been enlarged for more efficient detection of a proton.

\section{Summary}

We have constructed a new laser Compton backscattering beamline (LEPS2) at SPring-8, aiming 
one order of magnitude higher intensity and larger acceptance coverage compared with the LEPS 
experiments. We have successfully obtained the first photon beam at the LEPS2 beamline, 
resulting in the beam size of $\sigma \sim$6~mm and the intensity of $\sim$7~MHz for 
0$<$E$_\gamma$$<$2.4~GeV. The photon energy spectrum has been also measured using a large 
BGO crystal. 

We will start the BGOegg calorimeter experiments from the latter half of FY2013 with 
the forward coverage by TOF detectors and a DC. We will carry out $\eta$'-related physics 
programs, double meson photoproduction studies, etc. The momentum resolution less than 
1\% is expected for the proton which is detected by the RPC, and the $\eta$' mass 
resolution in the 2$\gamma$ decay mode is estimated to be $\sim$3\% without a kinematical 
fitting. Experiments with the LEPS2 charged particle spectrometer will follow the BGOegg 
experiments in a few years. Physics results at the LEPS experiments will be upgraded.

\section*{Acknowledgement}

We greatly thank to the support of the staff at SPring-8 for the beamline construction
and commissioning. This research was supported in part by the Ministry of Education,
Science, Sports and Culture of Japan, JSPS KAKENHI Grant No.~24244022, and Scientific 
Research on Innovative Areas Grant No.~21105003.

\end{multicols}


\begin{thebibliography}{}
\bibitem{muramatsu}     N.~Muramatsu, arXiv:1201.4094 (2012).
\bibitem{hwang}         S.H.~Hwang et al., Phys. Rev. Lett. 108, 092001 (2012).
\bibitem{morino}        Y.~Morino et al., Submitted Phys. Lett. B; 
                        Also Talked in BARYONS 2013, Hadron Spectroscopy Session 1-2.
\bibitem{clas}          M.~Williams et al., Phys. Rev. C 80, 045213 (2009).
\bibitem{yosoi}         M.~Yosoi, AIP Conf.~Proc.~1388 (2011) 163;
                        M.~Yosoi, SPring-8 Reasearch Frontiers 2011, 138.
\bibitem{yorita}        T.~Yorita, IPAC 2013 Proc.~(in press), WEPWA016.
\bibitem{ishikawa}      T.~Ishikawa et al., ELPH Annual Report Vol.1, 61 (2012).
\bibitem{nagahiro}      H.~Nagahiro et al., Phys. Rev. C 74, 045203 (2006).
\bibitem{tomida}        N.~Tomida, Master Thesis, Kyoto University (2012).
\bibitem{hashimoto}     T.~Hashimoto, Master Thesis, Kyoto University (2013).
\bibitem{fb20}          Y.~Kato, Few-Body Syst. (2013), DOI:10.1007/s00601-013-0681-6;
                        N.~Muramatsu, Few-Body Syst. (2013), DOI:10.1007/s00601-013-0614-4.
\end{thebibliography}
\end{document}